\documentclass[submission, Proceedings,a4paper]{SciPost}

\hypersetup{
    pdftitle={Light thermal relics enabled by a second Higgs},
    colorlinks,
    linkcolor={red!50!black},
    citecolor={blue!50!black},
    urlcolor={blue!80!black}
}

\usepackage[bitstream-charter]{mathdesign}
\newcommand{\sigmav}{\ensuremath{{\langle \sigma v \rangle}}}

\newcommand{\GeV}{\,\mathrm{GeV}}

\DeclareSymbolFont{usualmathcal}{OMS}{cmsy}{m}{n}
\DeclareSymbolFontAlphabet{\mathcal}{usualmathcal}

\begin{document}

\begin{center}{\Large \textbf{
Light thermal relics enabled by a second Higgs\\
}}\end{center}

\begin{center}
Johannes Herms\textsuperscript{1$\star$},
Sudip Jana\textsuperscript{1},
Vishnu P. K.\textsuperscript{2} and
Shaikh Saad\textsuperscript{3}

\end{center}

\begin{center}
{\bf 1} Max-Planck-Institut f{\"u}r Kernphysik, Saupfercheckweg 1, 69117 Heidelberg, Germany
\\
{\bf 2} Department of Physics, Oklahoma State University, Stillwater, OK, 74078, USA
\\
{\bf 3} Department of Physics, University of Basel, Klingelbergstrasse 82, CH-4056 Basel, Switzerland
\\
* herms@mpi-hd.mpg.de
\end{center}

\begin{center}
\today
\end{center}


\definecolor{palegray}{gray}{0.95}
\begin{center}
\colorbox{palegray}{
  \begin{tabular}{rr}
  \begin{minipage}{0.1\textwidth}
    \includegraphics[width=30mm]{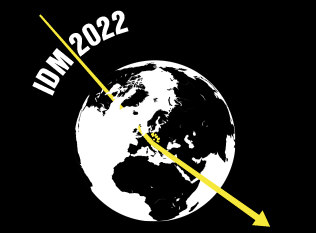}
  \end{minipage}
  &
  \begin{minipage}{0.85\textwidth}
    \begin{center}
    {\it 14th International Conference on Identification of Dark Matter}\\
    {\it Vienna, Austria, 18-22 July 2022} \\
    \doi{10.21468/SciPostPhysProc.?}\\
    \end{center}
  \end{minipage}
\end{tabular}
}
\end{center}

\section*{Abstract}
{\bf
Sub-GeV thermal relic dark matter typically requires the existence of a light mediator particle. We introduce the light two-Higgs-doublet portal, illustrated by a minimal UV-complete model for sub-GeV dark matter with kinematically forbidden annihilations into leptons.
All new physics states in this scenario lie at or below the electroweak scale, affecting Higgs physics, the muon anomalous magnetic moment and potentially neutrino masses. Observation of radiative dark matter annihilation by future MeV gamma-ray telescopes would be key to unambiguously identify the scenario.
}

\section{Introduction}
\label{sec:intro}
Dark matter (DM) is central to our understanding of the Universe, from the largest scales of cosmology down to the dynamics of individual galaxies.
Its constituents, however, are unknown.
Relic particles are a prime candidate to make up DM, and by studying mechanisms for their production in the early Universe, we can infer specific signatures to look for today.
Thermal relics that had sufficient interactions to enter into thermal equilibrium with the bath of Standard Model (SM) particles in the early Universe are particularly predictive, as thermalisation erases any prior cosmic history.

Traditionally, attention has been focused on electroweak-scale (EW) Weakly Interacting Massive Particles (WIMPs, see eg.~\cite{Bertone:2016nfn}), but thermal relic DM could be as light as a few MeV before the energy released in its freeze-out process disturbs successful Big Bang nucleosynthesis (eg.~\cite{Sabti:2019mhn}).
In this talk, we present a minimal and UV-complete realisation of light thermal DM~\cite{Herms:2022nhd}, based on a light scalar mediator that can arise in the well-known two-Higgs-doublet model (2HDM).

In the WIMP scenario, the DM relic density is determined by the freeze-out of annihilation reactions in the early Universe. The observed relic density $\Omega_\mathrm{DM}^\mathrm{obs} h^2 = 0.12$~\cite{Planck:2018vyg} is related to the thermally averaged annihilation cross section  
$\sigmav \sim \text{few} \times 10^{-9} \GeV^{-2}$.
This is easily achieved for EW-scale WIMPs, however there are two major challenges for sub-GeV WIMPs.

The first challenge is to accommodate the large needed couplings to the  MeV-scale Standard Model bath. Interactions mediated by the heavy particles of the SM are insufficient to appropriately deplete the DM before freeze-out (cf.\ Lee-Weinberg bound $m_\mathrm{DM}\gtrsim 2 \GeV$~\cite{Lee:1977ua} for $Z$-mediated DM annihilation; Higgs-mediated interactions do not fare any better).
New EW-scale particles coupled more strongly to the light SM degrees of freedom are experimentally disfavoured, leaving us with a new light mediator as the only option.

The second challenge for sub-GeV WIMPs stems from residual annihilation at late times. The annihilation rate scales as $n_\mathrm{DM}^2 \sigmav \propto m_\mathrm{DM}^{-2}$, and the associated energy release as $m_\mathrm{DM}^{-1}$. For DM masses $ m_\mathrm{DM} \lesssim 10 \GeV$ annihilation at late times must be suppressed compared to the value at freeze-out, $\sigmav_\mathrm{late} \ll \sigmav_\mathrm{fo}$, to not be in conflict with cosmic ray and CMB observations (eg.~\cite{Slatyer:2015jla}), unless its annihilation products are largely invisible, eg.\ neutrinos.

\section{The light 2HDM portal to dark matter}
In this section, we demonstrate how a light mediator can arise in the 2-Higgs-doublet model (2HDM). The 2HDM is a simple, well-studied extension of the SM, yet this possibility of very light scalars has been largely overlooked.
The model contains two scalar doublets $H_{1,2}$. We choose to work in the Higgs basis where only one of the neutral scalars obtains a nonzero vacuum expectation value (VEV):
\begin{align} 
	& H_1=\begin{pmatrix}
		G^{+}  \\
		\frac{1}{\sqrt{2}}(v+\phi_1^0+iG^0)    
	\end{pmatrix},\;    H_2=\begin{pmatrix}
		H^{+}  \\
		\frac{1}{\sqrt{2}}(\phi_2^0+iA)     
	\end{pmatrix}. 
\label{para}
\end{align}
Here $G^+$ and $G^0$ are the Goldstone modes, $H^+,\phi_1^0,\phi_2^0, A$ are the physical Higgs bosons, while the VEV $v\simeq 246$ GeV of $H_1$ governs EW symmetry breaking.
The scalar potential reads
\begin{align}
V&(H_1,H_2)= \; \mu_{1}^2H_1^{\dagger}H_1+\mu_{2}^2H_2^{\dagger}H_2
+\tfrac{\lambda_1}{2}(H_1^{\dagger}H_1)^2
+\tfrac{\lambda_2}{2}(H_2^{\dagger}H_2)^2 
+\lambda_3(H_1^{\dagger}H_1)(H_2^{\dagger}H_2)\\
&+\lambda_4(H_1^{\dagger}H_2)(H_2^{\dagger}H_1)
+\left\{
\big[
-\mu_{12}^2
+\tfrac{\lambda_5}{2}(H_1^{\dagger}H_2)
+\lambda_6(H_1^{\dagger}H_1)
+\lambda_7(H_2^{\dagger}H_2)
\big]
H_1^{\dagger}H_2+{\rm h.c.}\right\}
\,.\nonumber
\end{align}
We work in the alignment limit~\cite{Branco:2011iw,Babu:2018uik,Bernon:2015qea,BhupalDev:2014bir}, where $\phi_1^0\approx h$ is SM-like and decouples from the new CP-even Higgs $\phi_2^0\approx H$.
The masses of the scalar particles
in this limit
are
$m^2_h = \lambda_1v^2$,
$m^2_H =\mu^2_{2}+\frac{v^2}{2}(\lambda_3+\lambda_4+\lambda_5)$,
$m^2_A =m^2_H-v^2 \lambda_5$,
$m^2_{H^\pm} =m^2_H-v^2\frac{(\lambda_4+\lambda_5)}{2}$.
Clearly, $H$ can be made light independently from the other BSM scalars $A,H^\pm$ for appropriate values of $\lambda_{4,5}$. In the following, we consider
$m_H \ll m_A \sim m_{H^\pm}$.

Mass splitting between members of the same electroweak multiplet modify the electroweak oblique parameters, in particular the $T$-parameter~\cite{Grimus:2007if,Jana:2020pxx}. 
In the case of the 2HDM, the splitting between $H$ and $A$ on the one hand, and $H$ and $H^\pm$ on the other contribute with opposite sign. Values $m_{A,H^\pm} \lesssim 250 \GeV$ are accommodated easily, while much larger $m_{A,H^\pm}$ would require $A,H^\pm$ to be nearly degenerate.
Recently, the CDF collaboration~\cite{CDF:2022hxs} reported a measurement of the $W$-mass significantly larger than its SM expectation, which could indicate the existence of such a split multiplet~\cite{Babu:2022pdn,Chowdhury:2022moc}.

Direct searches at colliders constrain the scalar mass spectrum.
To forbid $Z\to H A$, we require $m_A>m_Z-m_H \sim 90\GeV$~\cite{Tanabashi:2018oca}. The production of a charged scalar in association with a $W$ boson at LEP requires $m_{H^\pm} \gtrsim 110\GeV$, while LHC searches for new charged scalars are insensitive for substantial $\mathrm{Br}_{\nu\tau}$ (see discussion in~\cite{Babu:2019mfe}, compare also~\cite{Iguro:2022tmr}).
The SM-like properties of the $h(125)$ observed at the LHC are largely unaffected by the second Higgs doublet in the alignment limit, except for contributions to invisible and radiative decay rates.
The SM-like Higgs may decay into the light new scalar, governed by the potential term $V \supset v h H^2 \frac{1}{2} \left(\lambda_3+\lambda_4+\lambda_5\right)$. Avoiding excessive exotic decays through this channel requires $\lambda_3 \simeq -(\lambda_4+\lambda_5) \sim m_{H^+}^2/v^2$.
The charged scalars contribute to the radiative decay rate into two photons. They interfere negatively with the dominant SM contribution, predicting a signal strength $R_{\gamma\gamma}<1$~\cite{Okawa:2020jea}.

To act as mediator, the new light scalar $H$ needs to couple to the light SM particles. This happens via the Yukawa couplings
$-\mathcal{L}_Y\supset  \widetilde{Y}_l \bar{\psi}_L H_1 \psi_R + {Y_l} \bar{\psi}_L H_2 \psi_R + \mathrm{h.c.}\,,$
where we focus on the coupling to leptons, for simplicity as well as possible connections to neutrino masses. 
In the alignment limit,  $\widetilde{Y}_l = \mathrm{diag}(m_e,m_\mu,m_\tau)/v$, while $Y_l$ is the mediator coupling responsible for DM phenomenology. 
The coupling to the light scalar affects the lepton anomalous magnetic moments and can explain the $(g-2)_{\mu}$ measurement at Fermilab~\cite{Muong-2:2021ojo,Jana:2020pxx}.
The off-diagonal entries of $Y_l$ are constrained by flavour violating processes, including anomalous $\tau$ and $Z$ decays (for details, see~\cite{Herms:2022nhd}).
A light scalar coupled to leptons can be produced at collider and beam dump experiments, as well as in supernovae, all of which constrain its couplings~\cite{Croon:2020lrf,Batell:2017kty,DAgnolo:2020mpt}.
When enlarging the model by a charged scalar $\eta^\pm$~\cite{Zee:1980ai}, the same Yukawa couplings responsible for the dark matter relic abundance can explain the observed neutrino masses~\cite{Herms:2022nhd}.

\section{Minimal sub-GeV WIMP dark matter}
To demonstrate the potential of the light 2HDM scalar as mediator between DM and the SM bath, we consider simple scalar and fermionic dark matter scenarios, each stabilised by a $\mathbb{Z}_2$ symmetry. In the fermionic scenario, the DM indirect detection issue of sub-GeV WIMPs is absent due to velocity suppression of annihilation processes.
In the scalar DM model, annihilation is kinematically forbidden at low temperatures.

\subsection{Minimal scalar forbidden dark matter}
The light 2HDM mediator can realise a minimal scenario for light forbidden dark matter.
We introduce a DM scalar $S$, stabilised by a dark parity symmetry. It couples to the SM via Higgs portal couplings
$\mathcal{L} \supset - \kappa_{ab} (S^\dagger S) (H_a^\dagger H_b)\,.$
The coupling $\kappa_{12}$ couples DM to the mediator $H$, which in turn can couple to the SM leptons.

Sub-GeV WIMP dark matter annihilating into charged particles is only viable if the late time annihilation rate is suppressed compared to its value at freeze out.
This is the case in the ``forbidden DM'' scenario~\cite{Griest:1990kh,DAgnolo:2015ujb,DAgnolo:2020mpt}, where there is a small mass splitting between the DM and its annihilation products, $\Delta=(m_{l_1}+m_{l_2}-2 m_\chi)/2 m_\chi>0$.
DM annihilation is then Boltzmann suppressed $ \sigmav_{\chi\chi\to l l } = \sigmav_{l l \to \chi \chi} e^{-2 \Delta (m_\chi/T)}$,
and kinematically forbidden for $T\to 0$.

We show relic abundance results in the scalar forbidden dark matter model coupled through the light second Higgs in Figure~\ref{fig:res} (fixing $\kappa_{ab} = 10^{-3}$ for concreteness). As a particularly interesting case, we consider the case of coupling exclusively to the muon (see~\cite{Herms:2022nhd} for more general results).
The blue (green) line shows the values of $Y_{\mu\mu}$ that reproduce the observed relic abundance for $\Delta=0$ ($\Delta=0.1$). The Boltzmann suppression caused by the larger mass splitting necessitates larger couplings to achieve the same relic density.
Laboratory and astrophysical constraints are shown in gray and purple. Parameters in the light green region can accommodate the $(g-2)_\mu$-value measured at Fermilab~\cite{Muong-2:2021ojo}.

Figure~\ref{fig:res} illustrates the power of DM indirect detection in identifying the present scenario.
For $\Delta \leq 0$, DM can annihilate directly into muons also at late times, in conflict with CMB observations (green shaded region).
We point out in~\cite{Herms:2022nhd} that radiative annihilation $SS \to \gamma\gamma$ is a powerful probe of forbidden DM annihilating into SM fermions. The green and blue shaded areas indicate upper bounds on $Y_{\mu\mu}$from the CMB for the corresponding mass splitting.
The dashed lines indicate the sensitivity of proposed medium energy $\gamma$-ray telescopes~\cite{AMEGO:2019gny,e-ASTROGAM:2017pxr,Engel:2022bgx} searching for dark matter annihilation at the Galactic center, which would almost certainly lead to discovery if the $Y_{\mu\mu}$-coupled scenario is realised in nature.

\subsection{Fermion dark matter}
\begin{figure}[hbt]
 \centering
 \includegraphics[width=0.45\textwidth]{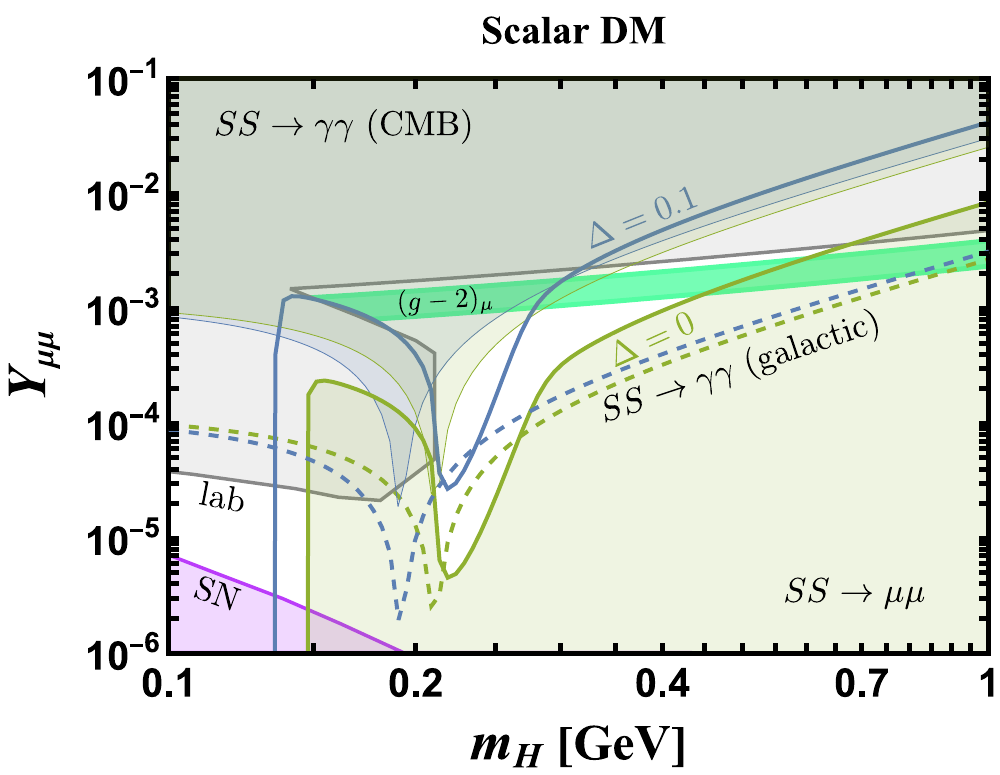}
 \includegraphics[width=0.45\textwidth]{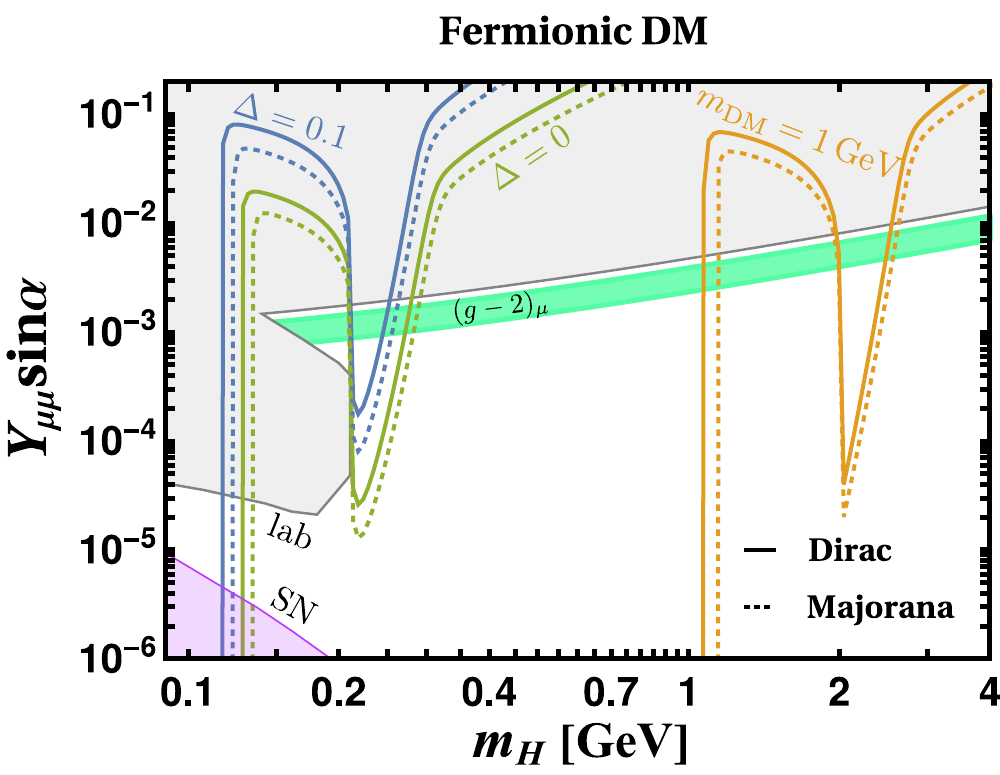}
 \caption{\emph{Left:} Forbidden scalar DM coupled to muons. Green and blue lines correspond to $\Delta=0,0.1$. Blue/green shaded areas show annihilation constraints on non-forbidden tree-level annihilation, as well as on radiative annihilation into photons, see text. Dashed lines indicate the sensitivity of proposed MeV $\gamma$-ray telescopes.
 \emph{Right:} Fermionic DM coupled to muons.
 The velocity suppressed annihilation requires larger couplings, while indirect detection bounds are absent.
 }
 \label{fig:res}
\end{figure}

A $\mathbb{Z}_2$-odd DM fermion $\chi$ cannot couple directly to the SM or the light scalar contained in the 2HDM in a renormalizable way.
As a result, we consider the presence of an additional real scalar $S$, even under $\mathbb{Z}_2$, which couples to the DM fermion through a Yukawa coupling $Y_\chi S \bar \chi\chi$.
The CP-even state contained in $H_2$ mixes with $S$, characterised by misalignment angle $\alpha$ that rotates $\phi_2^0$ and $S$ into the mass eigenstates $H$ and $H'$ (see~\cite{Herms:2022nhd} for details). Neglecting the contribution of $H'$ for simplicity, we recover a model very similar to the scalar scenario, where DM annihilation proceeds via $s$-channel exchange of the light mediator $H$ with a pair of SM fermions.
Results are shown in the right panel of Figure~\ref{fig:res} (taking $Y_\chi=0.1$). The annihilation process is velocity suppressed for both Dirac and Majorana DM, necessitating larger couplings than in the scalar scenario.
This restricts the model to the resonant region $m_H \gtrsim 2 m_\mathrm{DM}$, or alternatively to $m_H \sim m_\mathrm{DM}$ where forbidden annihilation into the mediator determines the relic abundance (vertical in plot).

\section{Conclusion}
The 2HDM contains the possibility of a sub-GeV scalar that can mediate between the dark sector and the SM through renormalizable couplings. All new physics in this scenario lies at or below the electroweak scale. Hints for the present scenario may already be 
showing, notably in the $(g-2)_\mu$ puzzle as well as the recent measurement of $m_W$ by CDF-II. We focused on a leptophilic second Higgs and leave the couplings to the light quarks to future work.

We demonstrate the capability of the light 2HDM portal in a fermionic and a scalar DM model. Annihilation in the fermionic case is velocity suppressed, while it is required to be kinematically forbidden in the scalar case, fixing the DM mass to be just below that of the SM leptons it annihilates to.
Galactic gamma ray lines at these energies from radiative DM annihilation would be a smoking-gun signature, observable at proposed sub-GeV $\gamma$-ray telescopes.

\paragraph{Funding information}
The work of VPK is in part supported by US Department of Energy Grant Number DE-SC 0016013.

\bibliography{reference.bib}

\nolinenumbers

\end{document}